\documentclass[sn-standardnature]{sn-jnl}

\usepackage[numbers]{natbib}

\newcommand{\mum}{\ifmmode{\rm \mu m}\else{$\mu$m }\fi}             
\newcommand{\Msun}{\ensuremath{{\rm M}_{\odot}}}                    

\newcommand{\chisq}{\ifmmode{\chi^{2} }\else{$\chi^2$}\fi}
\newcommand{\rchisq}{\ifmmode{\chi^{2} }\else{$\chi^2_\nu$}\fi}

\jyear{2022}%

\begin{document}

\title[The embedded young population of NGC~346]{Discovery of dusty sub-solar mass young stellar objects in NGC 346 with JWST/NIRCam}



\author*[1]{\fnm{Olivia C.} \sur{Jones}}\email{olivia.jones@stfc.ac.uk}

\author[2]{\fnm{Conor} \sur{Nally}}

\author[3]{\fnm{Nolan} \sur{Habel}}

\author[3]{\fnm{Laura} \sur{Lenki\'{c}}}

\author[4]{\fnm{Katja} \sur{Fahrion}}

\author[5]{\fnm{Alec S.} \sur{Hirschauer}}

\author[6]{\fnm{Laurie E.\ U.} \sur{Chu}}

\author[3]{\fnm{Margaret} \sur{Meixner}}

\author[7]{\fnm{Guido} \sur{De Marchi}}

\author[5]{\fnm{Omnarayani} \sur{Nayak}}

\author[5,8]{\fnm{Massimo} \sur{Robberto}}

\author[5]{\fnm{Elena} \sur{Sabbi}}

\author[9]{\fnm{Peter} \sur{Zeidler}}


\author[10]{\fnm{Catarina} \sur{Alves de Oliveira}}

\author[5]{\fnm{Tracy} \sur{Beck}}

\author[11]{\fnm{Katia} \sur{Biazzo}}

\author[12]{\fnm{Bernhard} \sur{Brandl}}

\author[8]{\fnm{Giovanna} \sur{Giardino}}

\author[13]{\fnm{Teresa} \sur{Jerabkova}}

\author[5]{\fnm{Charles} \sur{Keyes}}

\author[5]{\fnm{James} \sur{Muzerolle}}

\author[5]{\fnm{Nino} \sur{Panagia}}

\author[5]{\fnm{Klaus} \sur{Pontoppidan}}

\author[12]{\fnm{Ciaran} \sur{Rogers}}

\author[5,8]{\fnm{B.\ A.} \sur{Sargent}}

\author[5]{\fnm{David} \sur{Soderblom}}


\affil*[1]{\orgdiv{UK Astronomy Technology Centre}, \orgname{Royal Observatory}, \orgaddress{\street{Blackford Hill}, \city{Edinburgh}, \postcode{EH9 3HJ}, \country{UK}}}

\affil[2]{\orgdiv{Institute for Astronomy}, \orgname{University of Edinburgh}, \orgaddress{\street{Blackford Hill}, \city{Edinburgh}, \postcode{EH9 3HJ}, \country{UK}}}

\affil[3]{\orgdiv{Stratospheric Observatory for Infrared Astronomy}, \orgname{NASA Ames Research Center}, \orgaddress{\street{Mail Stop 204-14}, \city{Moffett Field}, \postcode{94035}, \state{CA}, \country{USA}}}

\affil[4]{\orgdiv{European Space Research and Technology Centre}, \orgname{European Space Agency}, \orgaddress{\street{Keplerlaan 1}, \city{Noordwijk},  \country{The Netherlands}}}

\affil[5]{\orgname{Space Telescope Science Institute}, \orgaddress{\street{3700 San Martin Drive}, \city{Baltimore}, \postcode{21218}, \state{MD}, \country{USA}}}

\affil[6]{\orgdiv{NASA Postdoctoral Program Fellow}, \orgname{NASA Ames Research Center}, \orgaddress{\street{M/S 245-1}, \city{Moffett Field}, \postcode{94035}, \state{CA}, \country{USA}}}


\affil[7]{\orgdiv{European Space Research and Technology Centre}, \orgname{European Space Agency}, \orgaddress{\street{Keplerlaan 1}, \city{Noordwijk}, \ \country{Netherlands}}}

\affil[8]{\orgname{Johns Hopkins University}, \orgaddress{\street{3400 N. Charles Street}, \city{Baltimore}, \postcode{MD 21218}, \country{USA}}}

\affil[9]{\orgdiv{AURA for the European Space Agency}, \orgname{Space Telescope Science Institute}, \orgaddress{\street{3700 San Martin Drive}, \city{Baltimore}, \postcode{21218}, \state{MD}, \country{USA}}}

\affil[10]{\orgdiv{ESAC}, \orgname{European Space Agency}, \orgaddress{\street{28692 Villafranca del Castillo}, \city{Madrid}, \country{Spain}}}

\affil[11]{\orgdiv{INAF}, \orgname{Astronomical Observatory of Rome}, \orgaddress{\street{Via Frascati 33}, \city{Monteporzio Catone}, \postcode{I-00078}, \country{Italy}}}

\affil[12]{\orgdiv{Leiden Observatory}, \orgname{Leiden University}, \orgaddress{\street{2300 RA Leiden}, \city{Leiden}, \postcode{PO Box 9513}, \country{The Netherlands}}}

\affil[13]{\orgname{European Southern Observatory}, \orgaddress{\street{ Karl-Schwarzschild-Strasse 2}, \city{Garching},\country{Germany}}}



\abstract{{\it JWST} observations of NGC 346, a star-forming region in the metal-poor Small Magellanic Cloud, reveal a substantial population of sub-solar mass young stellar objects (YSOs) with IR excess. We detected $\sim$500 YSOs and pre main sequence (PMS) stars from more than 45,000 unique sources utilizing all four NIRCam wide filters with deep, high-resolution imaging, where ongoing low-mass star formation is concentrated along dust filaments. From these observations, we construct detailed near-IR colour-magnitude diagrams with which preliminary categorizations of YSO classes are made. For the youngest, most deeply-embedded objects, {\em JWST}/NIRCam reaches over 10 magnitudes below {\em Spitzer} observations at comparable wavelengths, and two magnitudes fainter than {\em HST} for more-evolved PMS sources, corresponding to  $\sim$0.1~\Msun. For the first time in an extragalactic environment, we detect embedded low-mass star-formation. Furthermore, evidence of IR excess and accretion suggests that dust required for rocky planet formation is present at metallicities as low as 0.2 $Z_\odot$.}
 

\maketitle


\section{Introduction}
\label{sec:intro}

Located in the Small Magellanic Cloud (SMC) at a distance of $\sim$62\,kpc \citep{deGrijs2015}, NGC 346 is a prominent young cluster ($\sim$3 Myr; \cite{Bouret2003}) actively forming stars. It is the brightest and largest star-formation region in this metal-poor galaxy ($\sim$1/5 $Z_{\odot}$; \cite{Peimbert2000}) which has a comparable metallicity to galaxies at the epoch of peak star formation \citep[``cosmic noon";][]{Madau2014, Dimaratos2015}. 
Below these levels of chemical enrichment, the dust content of the interstellar medium (ISM) drops precipitously, altering the environment in which stars form (e.g., \cite{Tchernyshyov2015, RomanDuval2014}).

It is unknown if sufficient quantities of dust survives the star formation process in order to contribute to the formation of rocky planetary systems in low metallicity environments. 
Below a certain threshold of metal abundance, planetesimal formation via the streaming instability is suppressed \citep{Johansen09,Rixin21}.
The metallicity of inner protoplanetary disks is therefore thought to play a critical role in the ability to form terrestrial planets.
Further, the dust content of disks sets their lifetimes as lower-metallicity systems are more susceptible to fast photo-evaporation \citep{Ercolano10}. It is therefore of great interest to identify low-mass young stellar objects (YSOs) and discern their dust content.

The star formation history of NGC 346 is complex, with multiple stellar populations identified within the cluster \citep[e.g.,][]{Cignoni2011}. 
Powering the ionization of this giant H\,{\sc ii} region are more than 30 spectroscopically-identified massive (35--100 M$_\odot$) O-type stars \citep{Massey1989, Evans2006, Dufton2019}, the largest such sample in the SMC, which dominate the radiative and mechanical feedback.
Deep {\em Hubble Space Telescope} (HST) images reveal thousands of low-mass (0.6--3 M$_\odot$) pre main sequence (PMS) stars \citep{Nota2006}, which are distributed throughout the nebula and are connected by gas and dust filaments \citep{Sabbi2007, Hennekemper2008, DeMarchi2011b}. 
{\em Spitzer} and {\em Herschel} surveys of the SMC \citep{Bolatto2007, Gordon2011, Meixner2013} unveiled approximately 100 candidate YSOs in the very early stages of formation within the NGC 346 complex  \citep{Simon2007, Sewilo2013, Seale2014}. These high-mass YSOs possess typical masses of 8 M$_\odot$ and have formed within the past $\sim$1~Myr. They are located at the edge of or inside dusty pillars, which are often associated with H$\alpha$ emission. Their presence establishes that star formation is ongoing throughout the complex at a rate $>3.2 \times 10^{-3}$ ${\rm M}_{\odot} \, {\rm yr}^{-1}$ \citep{Simon2007}. 
In the infrared (IR), spectroscopic data for the young populations in NGC~346 are limited.
{\em Spitzer} IRS data spectroscopically confirmed the identity of six massive YSOs in the cluster \citep{Ruffle2015}.  \cite{Rubio2018} obtained HK band spectra to confirm the existence of three early-type stars in NGC~346.
Most recently, \cite{Jones2022} used VLT/KMOS to observe $\sim$15 other YSO candidates which were resolved into multiple young stars still accreting mass. 

Overall, NGC 346 possesses a complex distribution of hierarchically-linked star clusters of varying ages which inhabit a variety of environments \citep{Hony2015}, and which are dispersed across the extended field \citep{Sabbi2008, Hennekemper2008, Gouliermis2014}.
Within the ISM, there is a wide range of substructures exhibited in polycyclic aromatic hydrocarbon emission (PAH; 8 $\mu$m), warm dust (24 $\mu$m), and molecular gas \citep[CO J $= 2-1$;][]{Rubio2000, Contursi2000, Hony2015}.
A tight correlation is seen between the molecular gas, 8 $\mu$m emission and H$\alpha$, which presents as a well-defined bar extending from the centre of the region to the northeast and as an arc structure extending from southeast to northwest.
A recent Atacama Large Millimeter/submillimeter Array (ALMA) CO(J $= 1-0$) study \citep{Neelamkodan2021} discovered that the intersection of three colliding clumpy filaments is co-spatial with the locations of a cluster of YSOs and PMS stars.
Using {\em HST} proper motions and VLT/MUSE radial velocities,  \cite{Sabbi2022} and \cite{Zeidler2022} showed that stars in NGC~346 move along a wide spiral and that clusters of YSOs and young PMS stars seem to be predominately located where the coherent motion field shows significant changes, hence turbulence is still driving star formation across the system. 

NGC 346 is one of the most active star-forming regions in the Local Group. Its proximity, size ($\sim$100 $\times$ 100 pc$^2$), low foreground extinction, and an abundance of wide-field, high-resolution panchromatic data make it an ideal system for the study of both low- and high-mass star formation, the effects of this star formation on the surrounding medium, and the potential triggers of star formation in an environment vastly different from our local Galactic surroundings, and akin to galaxies at cosmic noon.

\section{Results and Discussion}
\label{sec:results}

\subsection{Images} 
\label{sec:images}

\begin{figure*}
\centering
\includegraphics[trim=0cm 0cm 0cm 0cm, clip=true,width=\textwidth]{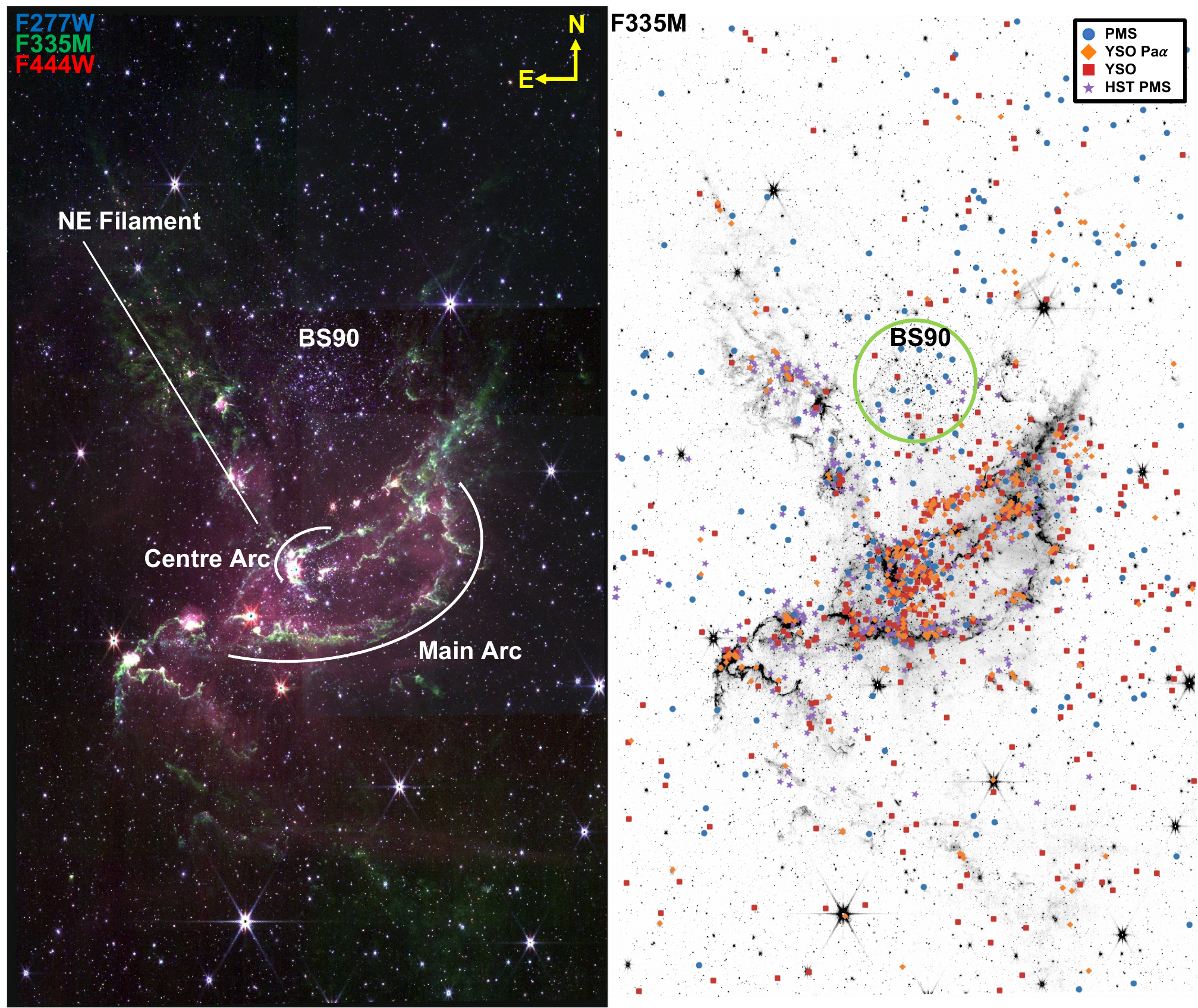}
\caption{NIRCam mosaics of NGC 346.
 Left: Three-colour composite mosaic of NGC 346 combining the F277W (blue), F335M (green), and F444W (red) filters. The region is rich in structures of knots, arcs, and filaments.
 Areas of bright pink/red emission are associated with clumpy star formation. The spatially-resolved PAH emission excited by UV photons in green is brightest in regions corresponding to the edges of dense material, characteristic of a photodissociation region (PDR). Massive stars, stars belonging to the BS90 cluster (circled in green), and the SMC field population are also visible. Right: Mosaic image from the F335M filter showing four different populations of stars (see Fig. \ref{fig:NGC346_CMDs}): PMS stars (blue circles), YSOs with Pa{$\alpha$} (orange diamonds), YSOs without Pa{$\alpha$} (red squares), and stars matched with the HST PMS catalogue (purple stars).   
} 
 \label{fig:NGC346_JWST_3colour}
 \end{figure*}

We observed NGC 346 with {\em JWST}, using the Near Infrared Camera (NIRCam; \cite{Rieke2005}) F115W, F187N, F200W, F277W, F335M, and F444W filters.
The NIRCam images of NGC 346 shown in Figure~\ref{fig:NGC346_JWST_3colour} reveal the complex filamentary structure of the NGC 346 main arc, dominated by emission from warm dust and PAHs, together with the intermediate-age BS90 cluster \citep{Bica1995} just above the centre. 
The brightest red stars are located along dust ridges, in tips of warm dust lanes, in large sub-clusters within the centre NGC 346 arc, or in smaller clumps located along the main arc and northeast perpendicular filament. 
The images show large variations in the dust morphology and highlight feedback from the complex star formation history of the region.
The impact of star formation and stellar feedback is revealed by the heating of the dust and fluorescing PAHs due to C-H bond stretching in the F335M band \citep{Rodriguez2022, Dale2022, Sandstrom2023}.
This occurs on the edges of the main arc structure illuminated by UV photons from massive stars compared to the surrounding T$\sim$ 600~K dust seen in the F444W band. 
The northeast filament perpendicular to the NGC 346 main body  extends further than what is seen in {\em HST} data and is brightest in the F335M band. 

\subsection{Identification of Young Stellar Objects}
\label{YSO}

\begin{figure*}
\centering

\includegraphics[trim=0cm 0cm 0cm 0cm, clip=true,width=0.4\textwidth]{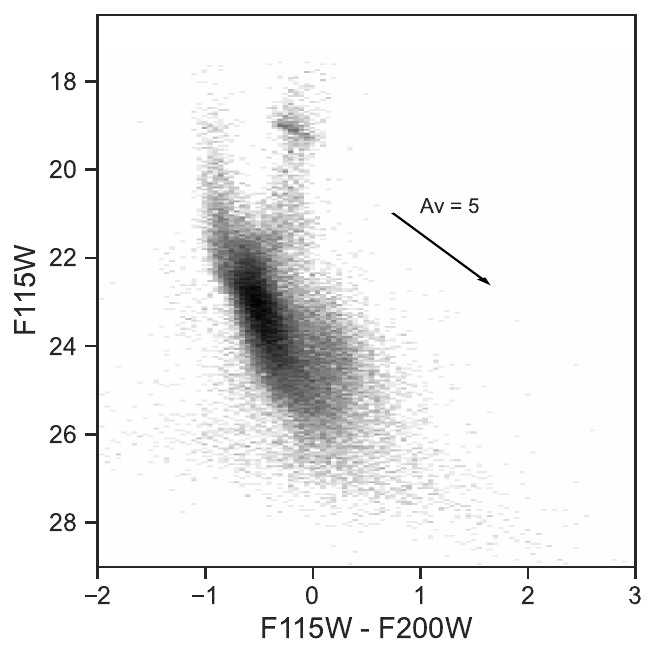}
\includegraphics[trim=0cm 0cm 0cm 0cm, clip=true,width=0.4\textwidth]{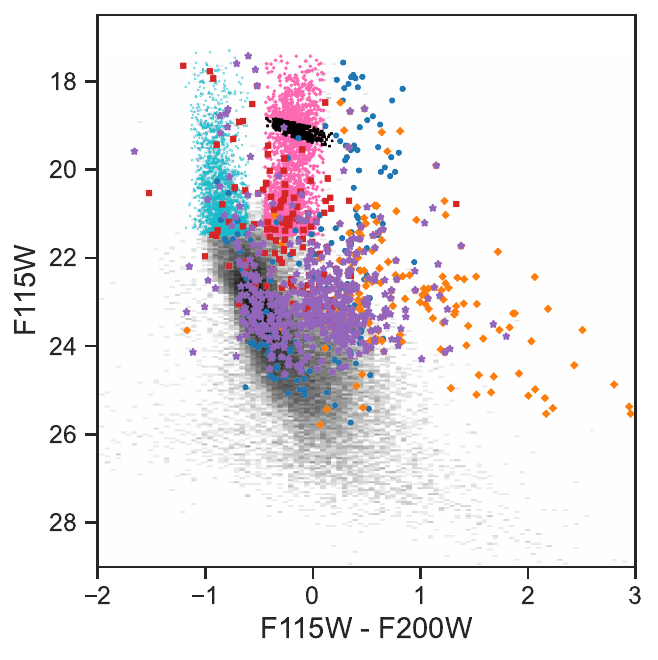}
\includegraphics[trim=0cm 0cm 0cm 0cm, clip=true,width=0.4\textwidth]{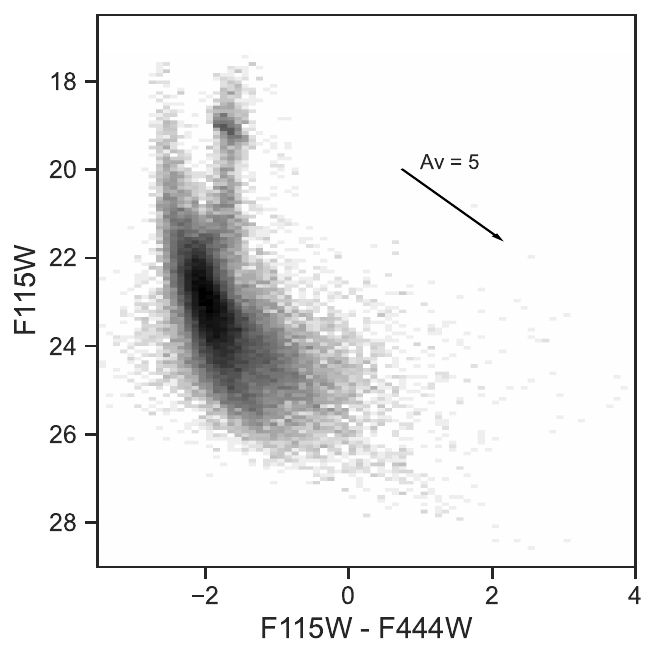}
\includegraphics[trim=0cm 0cm 0cm 0cm, clip=true,width=0.4\textwidth]{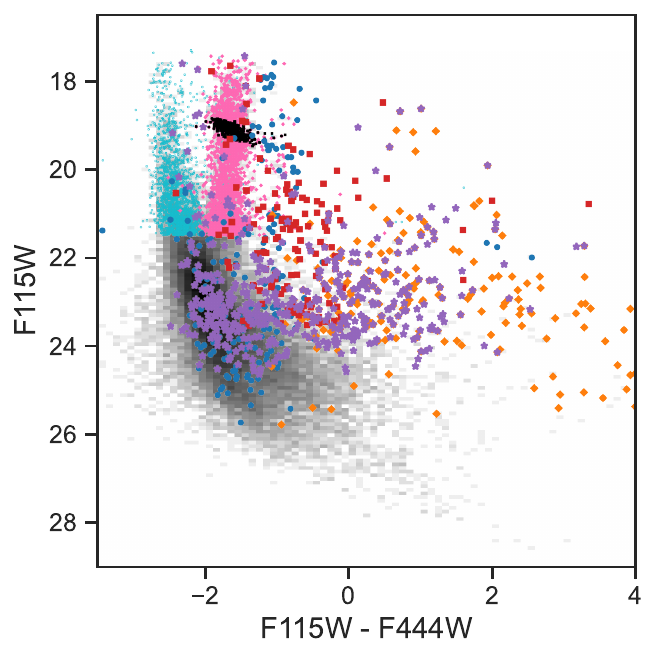}
\includegraphics[trim=0cm 0cm 0cm 0cm, clip=true,width=0.4\textwidth]{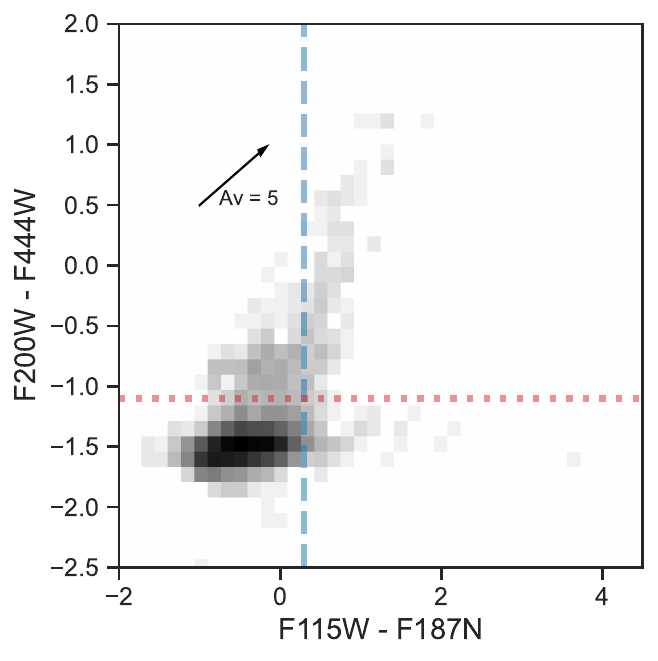}
\includegraphics[trim=0cm 0cm 0cm 0cm, clip=true,width=0.4\textwidth]{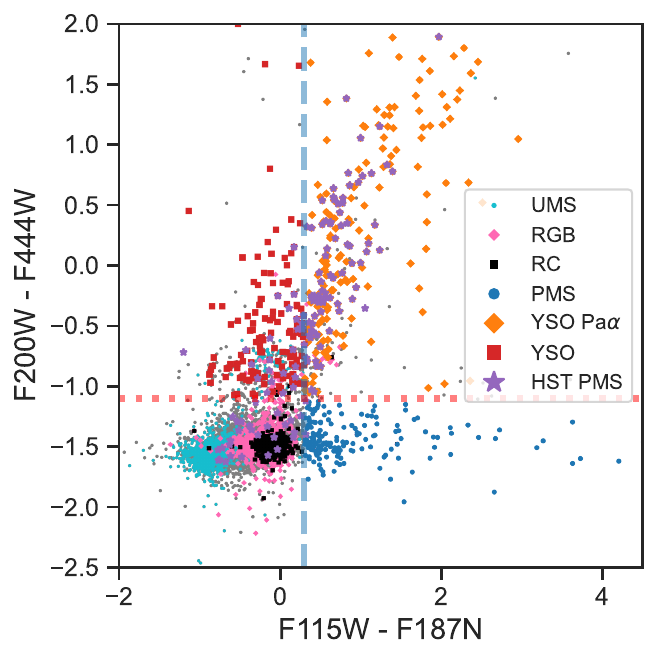}
 \caption{Colour-magnitude diagrams and colour-colour diagrams highlighting sources classified photometrically (right) and as stellar density Hess diagrams (left) for all point sources in the NGC 346 catalogue. 
 The red giant branch (RGB), red clump (RC), main sequence turn-off (MSTO), and upper main sequence (UMS) are all easily identified in the stellar populations. 
 In the bottom panels, the red dotted line separates sources with an IR excess, whilst the blue dashed line separates objects with a Pa{$\alpha$} excess from the rest of the population.} 
 \label{fig:NGC346_CMDs}
 \end{figure*}

As they are enshrouded in collapsing dusty envelopes and accretion disks \citep{Lada1987, Robitaille2006, Whitney2008}, young YSOs are best identified utilising IR colours. As they evolve and the circumstellar envelopes and disks dissipate, the central star becomes more apparent in shorter-wavelength light. 
From our initial band-matched photometric catalogue,  45,583 unique point sources were detected in all four of our wide-band NIRCam filters.
This selection provides the most reliable colour-magnitude diagrams (CMDs) at the cost of some additional photometric depth. 

CMDs constructed from the galactic extinction-corrected photometry are presented in Figure~\ref{fig:NGC346_CMDs}. The details of the aperture photometry and corrections are given in Section~\ref{sec:Photometry}. In these near-IR CMDs, the populations of red giant branch (RGB), red clump (RC), and upper main sequence (UMS) stars are clearly separated from the dominant population of lower main-sequence and PMS stars. Table~\ref{tab:colourSelection} lists the colour selection criteria for these populations and the number of sources in each class.  Only {\em JWST}/NIRCam data was used in these classifications.
Evolved stars (e.g., red supergiants, asymptotic giant branch (AGB), and post-AGB stars) are brighter than the saturation limit and thus not expected to appear in this CMD. Furthermore, as NIRCam (PSF $\sim 0.1$~arcsec) resolves structures for distant galaxies, contamination from background galaxies and extended sources is negligible in our point-source catalogue and corresponding CMDs. 

The F115W filter is essential to the identification of UMS, RC, and RGB stars, as these sequences, as well as main sequence turn-off (MSTO) stars, are conflated in CMDs utilising longer-wavelength colours. In our CMDs, elevated photometric uncertainties cause scatter for sources near our detection limit, while a spread of ages and differential extinction across the field broadens the shape of the evolutionary sequences at all magnitudes. To visualize the effect of extinction, we show on the diagrams the reddening vector corresponding to $A_V$=5  according to the SMC Bar Average Extinction Curve of \cite{2003ApJ...594..279G}.  

When compared to optical CMDs derived from {\em HST} data, we find more than 6,000 sources that are consistent with PMS stars in the mass range between $\sim 0.5$ and $4\, {M_\odot}$ based on their colours and magnitudes \citep{Nota2006, Sabbi2007}.  
Furthermore, we find candidate PMS stars in F115W extending at least two magnitudes below the {\em HST} detection limit, suggesting that we can observe T-Tauri stars down to $\sim 0.1\, M_\odot$. 
\cite{DeMarchi2011b} used H$\alpha$ excess to identify bona fide PMS with active accretion. Of these PMS candidates, 435 have a match in the NIRCam point source catalogue, we refer to these sources as HST-PMS in the figures.

In order to disentangle young stars ($<$10~Myr) from other populations in the field, we use an F200W--F444W versus F115W--F187N colour-colour diagram (2CD; Figure~\ref{fig:NGC346_CMDs}). Narrow-band photometry with the F187N filter traces the hydrogen Pa{$\alpha$} recombination line at 1.875 $\mu$m, characteristic of young PMS stars undergoing mass accretion. Approximately 8500 sources are reliably detected in F187N. When this can be unambiguously identified, the accretion luminosity and mass accretion rate can be derived following methods similar to that developed by \cite{DeMarchi2011a} using H$\alpha$. 

The F200W--F444W versus F115W--F187N 2CD shows a substantial concentration of sources, with the UMS clustered around the ($-0.6, -1.5$) point and the RC and RGB stars slightly to the right at ($0, -1.5$). To conservatively identify the YSOs, we define a horizontal line (F200W--F444W$=-1.1$) and a vertical line (F115W--F187N$=0.3$) that enclose almost all of the UMS, RC, and RGB stars. In particular, the vertical line ensures that we are not including RGB stars with winds that may have a Pa$\alpha$ excess. 

A minority of sources appear to spread in the upper-right direction, roughly following our representative SMC reddening vector. In principle, these sources could be interpreted as either highly-reddened objects due to their circumstellar material or as objects with substantial IR excess, possibly associated with accretion emission in the F187N filter. 

To break this degeneracy, we may look at the distribution of these outliers in the 2CD (see Sec.~\ref{sec:location}). Examining the quadrants defined with respect to ($0.3, -1.1$), objects in the bottom-left quadrant are generally compatible with the main populations, with negligible reddening.
Second, objects in the bottom-right quadrant (blue dots) have F200W--F444W compatible with stellar photospheres, but show significant F187N excess. One of these sources also showed H$\alpha$ excess when observed with {\em HST}. Strong line emission in this case is the most viable explanation, with mass accretion, possibly a sporadic large episode, as a plausible source. In this case, the lack of F444W excess may suggest that the accreting disk has cleared its inner hot-dust component and accretion is supported largely by the gaseous phase. The spatial location of these objects in the NGC 346 field, several of them rather bright in the F115W filter, shows some concentration in correspondence with the brightest clumps of nebular emission.  A significant fraction is spread in the field, however, suggesting that these accreting YSOs may be relatively evolved and dispersed. For simplicity, we refer to these objects as PMS objects.

Third, objects in the top-left quadrant (red squares), which we shall refer to as YSOs, have near-IR colours compatible with stellar photospheres and significant F444W excess, a characteristic incompatible with reddened objects. Of these, 29 also showed an H$\alpha$ excess in the {\em HST} catalogue. These objects are clustered at the centre of the region but become more spread out in the southern part of our field. This latter grouping may be a candidate for transitional YSOs (i.e., protoplanetary disks with inner holes in the dust distribution and negligible mass accretion).

Finally, the last class of objects (orange diamonds) is comprised of sources in the top-right quadrant of the 2CD. In the F115W--F444W CMD, they appear well above the region occupied by low-mass MS and PMS stars, suggesting that they may be highly-reddened, relatively massive stars. {\em HST} photometry indicates that 80 of these objects also show H$\alpha$ excess. 
Their spatial distribution strikingly traces the main filaments of the region, suggesting that they are associated with ongoing star-formation sites. Remarkably, a large fraction of the accreting PMS stars detected by {\em HST} lie in this sector, supporting the hypothesis that these are bona fide YSOs that have not yet significantly migrated from their birthplace. 
We shall refer to them as YSOs with Pa$\alpha$ emission.

This simplified, yet conservative selection of YSO and PMS candidates represents the deepest census of a star-forming region in a low-metallicity galaxy. Our data include redder candidates that are not present in {\em HST} optical catalogues because they are undetectable in those bands, as well as low-mass ($\lesssim$2~\Msun) sources significantly below the completeness limit of {\em Spitzer} surveys. Typically YSOs in the early stages of formation can be completely obscured at optical wavelengths and faint in the near-IR,
however the unprecedented sensitivity of NIRCam enables the detection of sources with colours that are consistent with embedded low-mass YSOs; although additional mid-IR data is necessary to assess the physical properties of the most embedded YSOs with ages $<$ 0.1 Myr and to identify Class 0 sources \citep[e.g.,][]{Jones2017a}.

\begin{table*}
    \small
	\centering
     \setlength\tabcolsep{3.5pt} 
	\caption{NGC 346 Stellar Populations identified using {\em JWST}/NIRCam.}
	\label{tab:colourSelection}
	\begin{tabular}{llc} 
		\hline
		Population & Colour Selection & Number of Sources  \\
		\hline
		RC        & inside[x1,y1=(-0.45,18.8), x2,y2=(-0.45, 19.10),            &   448   \\
 		           &  ~~~~~~~ x3,y3=(0.10,19.55), x4,y4=(0.10,19.15)]        &    \\
 &   where x = F115W \& y=F115W--F200W  & \\
		RGB       &  0.1 $>$F115W--F200W $>$ $-$0.45 and F115W $< 21.5$                             &  2176   \\
		UMS       &  $-$0.6 $>$F115W--F200W $>$ $-$1.18 and F115W $< 21.5$                            &  1982   \\
		\hline
	YSO              & F115W$-$F187N $<$ 0.3 and  F200W$-$F444W $>$ $-$1.1  &    136  \\
 &  and F335M$-$F444W $>$ $-$0.3  &  \\
	YSO Pa{$\alpha$}   & F115W$-$F187N $>$ 0.3 and  F200W$-$F444W $>$ $-$1.1  &    216  \\
 &  and F335M$-$F444W $>$ $-$0.3  &  \\ 
	PMS              & F115W$-$F187N $>$ 0.3 and  F200W$-$F444W $<$ $-$1.1                            &    179  \\		
		\hline
	\end{tabular}
\end{table*}


\subsection{Spatial Distribution of NGC 346 stars} 
\label{sec:location}

Figure~\ref{fig:NGC346_JWST_3colour} shows there is a high degree of spatial overlap between YSOs (red squares), YSOs with Pa{$\alpha$} excess (orange diamonds), and the bright F444W dust emission.
Interestingly, at optical wavelengths, many of these sources are either not visible ($\sim27\%$) or 
 may be erroneously misidentified as main-sequence stars (de Marchi in prep.), highlighting the importance of IR observations to accurately interpret star-forming regions. We note that the location of both YSOs and YSOs with Pa{$\alpha$} excess do not necessarily correspond to the clusters previously identified in the {\em HST} data \citep{Sabbi2007}. On the contrary, they tend to encircle the cavities created by the NGC~346 OB stars \citep{Zeidler2022}. The distribution of the PMS stars and the newly-identified YSOs supports the scenario proposed by \cite{Sabbi2022}, in which a global hierarchical collapse culminates in ``river-like'' structures responsible for the formation of clumps where significant changes in the coherence of the motion field are detected, and therefore where one expects high gas friction.

On the other hand, the older PMS stars (blue dots) which have cleared their dust envelopes and are optically thin are diffusely distributed across the field. This larger spatial dispersion is consistent with formation episodes over the past $20$~Myr \citep{DeMarchi2011c, DeMarchi2011b}, and is likely due to both turbulent star formation and early dynamical evolution, in agreement with the spiralling nature and increasing rotation with distance from the centre of NGC~346 \citep[e.g.,][]{DeMarchi2011c, Cignoni2011, Sabbi2022, Zeidler2022} due to hierarchical collapse \citep{VazquezSemadeni2019}.



As expected, we find that {\em JWST} surpasses the capability of {\em Spitzer} to detect candidate YSOs using only aperture photometry. This expands the observed sample of approximately 100 YSO candidates within NGC 346 by over a factor of three. Our survey reveals a population of dusty, sub-solar mass YSOs, and represents the deepest extragalactic census of these objects at low metallicity. The discovery of an associated IR excess in these objects reveals for the first time the dust around low mass YSOs in a 0.2 $Z_\odot$ extra-galactic environment, suggesting that the material to form rocky planets is present at this low metallicity.

\section{Methods}
\subsection{NIRCam Observations and Data Processing}
\label{sec:observations}

We have mapped NGC 346 with {\em JWST}/NIRCam (Program ID:\ 1227; PI:\ Meixner), in the F115W, F187N, and F200W short-wavelength (SW) bands, and F277W, F335M, and F444W long-wavelength (LW) bands. 
The images, obtained on 2022 July 16 are centred at R.A.\ = 00:59:04.9451, decl.\ = $-$72:10:9.15, and cover an area of $\sim$31.05 arcmin$^2$ (see Table \ref{tab:obs_sum}). The NIRCam observations employed both the A and B modules to provide the largest field of view with one pointing. They were obtained in a four-by-four mosaic using the {\sc bright2} readout pattern with two groups per integration at four sub-pixel dither positions for an exposure time of 171.8 seconds per filter, for three of the tiles. The last tile was observed with seven groups per integration, to enable MIRI parallels, for a total exposure time of 601.3 seconds per filter. 

The level two NIRCam data were reprocessed using a slightly modified version of the {\em JWST} official pipeline (version 1.7.2).
These modifications correct for 1/f noise (using \texttt{image1overf.py} \citep{Willott_1f_2022}), flat field correction noise, World Coordinate System (WCS) alignment issues, differences in background matching across the mosaic, and include the most recent NIRCam calibration files \texttt{jwst\_0989.map} of the Operational Pipeline Calibration Reference Data System produced on 2022-10-03 with on-sky derived photometric zero-points \citep{Gordon2022, Boyer2022}.
The final pixel scale of the mosaics is set to  0.''0315 for the three SW bands and 0.''0629 for the three LW bands.  

\begin{table*}
    \setlength\tabcolsep{3.5pt} 
	\centering
    \small
	\caption{Summary of the NGC 346 NIRCam survey, Guaranteed Time Program \#1227 and values adopted for the properties of NGC 346. The completeness limit corresponds to aperture photometry with a 1.5-pixel radius to 3$\sigma$.}
	\label{tab:obs_sum}
	\begin{tabular}{lc} 
		\hline
		Characteristic & Value \\
		\hline
		Nominal center point              & 00:59:04.9451 $-$72:10:9.15\\
		Survey area (arcmin$^2$)          & 31.05 \\
		Central $\lambda$ ($\mu$m)        & 1.154, 1.874, 1.990,    2.786, 3.365, 4.421 \\
		FWHM at $\lambda$ (pixel)        & 1.290, 2.065,  2.129,   1.460,  1.762, 2.302 \\
		Point source completeness limits at $\lambda$ (mag)  & 26.6, 23.2, 26.7, 25.4, 24.5, 26.4 \\
        Maximum total exposure time per pixel        & 601.3 seconds \\  
		\hline
		Distance to NGC 346                & 60.4 kpc   \\
		Distance modulus (m $-$ M)$_0$     & 18.96    \citep{deGrijs2015}   \\
		$E(B - V)$                        &  0.08       \\
		Metallicity  [Fe/H] (dex)          &  $-$0.9--1.0   Z =\,0.002 \\
		\hline
	\end{tabular}
\end{table*}

\subsection{Photometry}
\label{sec:Photometry}

Aperture photometry was performed on the individual exposures in each band using the {\sc starbug ii} tool \citep{Nally_Starbug2_2022}. {\sc starbug ii}, which incorporates modules from {\sc photutils} \citep{photutils}, is optimized for observations utilizing both NIRCam and the Mid-Infrared Instrument \citep[MIRI;][]{Rieke15} on {\em JWST} and is designed to detect and extract point sources in crowded environments with complex diffuse emission and variable backgrounds.
The sources identified in the single frames are extracted at a 3$\sigma$ level above the local background, which was characterized and globally subtracted using a combination of three different background estimation techniques. This ensures objects in complex nebular regions in which background determination is more problematic are not prematurely excluded.
An aperture with radius 1.5 pixels and an annulus from 3.0 to 4.5 pixels surrounding each source was then employed in the photometric extraction.
Sharp between 0.4-0.9 and round $\leq\ |1.0|$ cuts are applied, and then only sources detected in at least three of the four frames are retained.
Sources with mean and median values that differ by more than 0.1 dex between exposures were flagged and removed as mismatches.
This eliminates cosmic rays, noise spikes from the point spread function (PSF), and extended sources such as resolved background galaxies, and ensures high fidelity of the final point source catalogues. 
Aperture corrections provided in the CRDS reference files were then applied to all photometry.
To generate a band‐merged point‐source catalogue, the individual catalogues from each of the six filters were merged using the closest astrometric separation $<$0''.25. Source blending/confusion is negligible ($<$3\%) when combining between different filters. We correct the photometric values for Galactic foreground extinction using $E(B - V) = 0.08$ and the extinction curve of \cite{Cardelli1989} with $R_{\rm V}$ = 2.7, but not for any extinction intrinsic to NGC 346.  
The final band-matched catalogue includes $\sim$525,000 unique point sources, of which 45,583 are reliably detected in all four NIRCam wide bands. We refer to the latter sample as our NIRCam source catalogue, which we present in AB magnitudes.
Point-sources in this catalogue have fewer nulled wavelengths, smaller photometric uncertainties and are typically brighter, thus less likely to have mismatched photometry when cross-matching with other telescopes (e.g., {\em HST, Gaia}).

In the F115W band, the point source completeness magnitude of 26.6 allows for the characterization of young populations ($<$10 Myr) down to an initial mass of $\sim$ 0.1 \Msun, corresponding to stars in the T-Tauri range. Sources brighter than F115W $=$ 17.3 mag are saturated. To verify the PMS mass limits, we match (using  $R < 0.3''$) our NIRCam catalogue to the \cite{Sabbi2008} and \cite{DeMarchi2011b} {\em HST} data which include mass and age estimates. There are 24,367 sources in common, including PMS stars with masses from 0.4--4 \Msun and ages 1--30 Myr.







\def\aj{AJ}					
\def\actaa{Acta Astron.}                        
\def\araa{ARA\&A}				
\def\apj{ApJ}					
\def\apjl{ApJL}					
\def\apjs{ApJS}					
\def\ao{Appl.~Opt.}				
\def\apss{Ap\&SS}				
\def\aap{A\&A}					
\def\aapr{A\&A~Rev.}				
\def\aaps{A\&AS}				
\def\azh{AZh}					
\def\baas{BAAS}					
\def\jrasc{JRASC}				
\def\memras{MmRAS}				
\def\mnras{MNRAS}				
\def\pra{Phys.~Rev.~A}				
\def\prb{Phys.~Rev.~B}				
\def\prc{Phys.~Rev.~C}				
\def\prd{Phys.~Rev.~D}				
\def\pre{Phys.~Rev.~E}				
\def\prl{Phys.~Rev.~Lett.}			
\def\pasp{PASP}					
\def\pasj{PASJ}					
\def\qjras{QJRAS}				
\def\skytel{S\&T}				
\def\solphys{Sol.~Phys.}			
\def\sovast{Soviet~Ast.}			
\def\ssr{Space~Sci.~Rev.}			
\def\zap{ZAp}					
\def\nat{Nature}				
\def\iaucirc{IAU~Circ.}				
\def\aplett{Astrophys.~Lett.}			
\def\apspr{Astrophys.~Space~Phys.~Res.}		
\def\bain{Bull.~Astron.~Inst.~Netherlands}	
\def\fcp{Fund.~Cosmic~Phys.}			
\def\gca{Geochim.~Cosmochim.~Acta}		
\def\grl{Geophys.~Res.~Lett.}			
\def\jcp{J.~Chem.~Phys.}			
\def\jgr{J.~Geophys.~Res.}			
\def\jqsrt{J.~Quant.~Spec.~Radiat.~Transf.}	
\def\memsai{Mem.~Soc.~Astron.~Italiana}		
\def\nphysa{Nucl.~Phys.~A}			
\def\physrep{Phys.~Rep.}			
\def\physscr{Phys.~Scr}				
\def\planss{Planet.~Space~Sci.}			
\def\procspie{Proc.~SPIE}			
\let\astap=\aap
\let\apjlett=\apjl
\let\apjsupp=\apjs
\let\applopt=\ao


\bibliography{main}{}

\backmatter





\bmhead{Code Availability Statement}
This research made use of astropy \citep{Astropy2013},  photutils \citep{photutils} and {\sc topcat} \citep{Taylor2005}. The {\sc starbug ii} tool \citep{Nally_Starbug2_2022} optimized for {\em JWST} NIRCam and MIRI point source photometry in complex crowded environments is available via {\sc pip install starbug2}.

\bmhead{Data Availability}


The data that support the findings of this study are available from the corresponding author upon reasonable request.

\bmhead{Competing Interests}

The authors declare no competing interests.

\bmhead{Author Contributions}

OCJ led the analysis and is the science lead of the NGC 346 Team. 
CN produced the photometric catalogues.
NH and LL reprocessed the NIRCam data.
KF and CR assisted in the photometry. 
MR, GdM, ES provided advice on NIRCam data processing and the analysis on comparison to {\em HST} data. 
LC produced images on NGC 346.
AH, MM, KP helped optimise the observations. 
All authors contributed to observation planning and/or scientific interpretation.

\bmhead{Acknowledgments}

This work is based on observations made with the NASA/ESA/CSA James Webb Space Telescope. The data were obtained from the Mikulski Archive for Space Telescopes at the Space Telescope Science Institute, which is operated by the Association of Universities for Research in Astronomy, Inc., under NASA contract NAS 5-03127 for JWST. These observations are associated with program \#1227.
OCJ acknowledge support from an STFC Webb fellowship.  
KF acknowledges support through the ESA Research Fellowship.
MM and NH acknowledge support through a NASA/JWST grant 80NSSC22K0025, and MM and LL acknowledge support from the NSF through grant 2054178.
ON acknowledges support from STScI Director's Discretionary Fund.

\end{document}